\documentstyle[amsfonts,amscd,verbatim,hyperref]{amsart}

\theoremstyle{plain}
\newtheorem{Thm}{Theorem}
\newtheorem{Prop}{Proposition}%[section]
\newtheorem{Lem}{Lemma}
\newtheorem{Cor}{Corollary}

\theoremstyle{remark}

\newcommand\real{\Bbb R}
\newcommand\escript{{\cal E}}
\newcommand\brac[1]{\left\{#1\right\}}
\newcommand\elbows[1]{(#1)}
\newcommand\hscript{{\cal H}}

\newcommand\pscript{{\mathcal P}}
\newcommand\wscript{{\mathcal W}}
\newcommand\half{\frac{1}{2}}
\newcommand\ab[1]{\left|#1\right|}  

\newcommand\sqbrac[1]{\left[#1\right]}  
\newcommand\paren[1]{\left(#1\right)}

\begin{document}
\author{P.~ Busch}
\address{\rm Department of Mathematics, The University of Hull}
\email{P.Busch@@maths.hull.ac.uk}
\author{S.P.~ Gudder}
\address{\rm Department of Mathematics and Computer Science, 
University of Denver}
\email{sgudder@@cs.du.edu}
\title{Effects as Functions
on Projective Hilbert Space}
\date{Published in Lett.~Math.~Phys. {\bf 47}, 329-337 (1999), 
\href{http://dx.doi.org/10.1023/A:1007573216122}{DOI: 10.1023/A:1007573216122}}
\thanks{Dedicated to Professor G.~ Ludwig.
The quest for a proof of Ludwig's theorem was brought to our
attention by Pekka Lahti.}  
\maketitle

\begin{abstract}
The set of effect operators in a complex Hilbert space can be
injectively embedded into the set of functions from the set of
one-dimensional projections to the real interval [0,1]. Properties of
this injection are investigated.
\end{abstract}

\section{Introduction}

In his monumental treatise, {\sl Foundations of Quantum Mechanics}, G.~
Ludwig formulates an intriguing characterization of Hilbert space
effects: any effect $g$ is uniquely ``determined by the maximal
$\overline\lambda$ for which $g\ge\overline\lambda P_{\varphi}$ for all
$\varphi$" (\cite{Lud}, p.\ 228). Here $P_{\varphi}$ denotes the
projection onto the subspace spanned by the nonzero vector $\varphi$. As
far as we can see, he sketches the proof by considering the case of
finite rank effects. Here we provide two distinct proofs for the general
case, one more order theoretic, the other more analytic in
flavour and each bringing out different aspects of this characterization
of effects.

\section{The Strength of an Effect Along a Ray}

Let ${\cal E} ({\cal H})$ be the set of effects on a Hilbert space
$\hscript$, that is, the set of positive operators dominated by the
identity operator $\mathrm{I}$. 
Throughout the paper, $\varphi$ will denote a
unit vector in $\hscript$. For $E\in{\cal E} (\hscript)$, $\varphi\in
\hscript$ with  $\|\varphi\|=1$, let 
 \begin{equation}
\lambda (E,P_{\varphi} )  
:=\sup\left\{\lambda\in{{\Bbb R}\colon\ \lambda P_\varphi\le E}\right\}.
\label{lam}
\end{equation}
It is clear that $0\le\lambda (E,P_{\varphi} )\le 1$.

We will call $\lambda (E,P_\varphi )$ the {\sl strength of} 
$E$ {\sl along} $P_\varphi$ and  $\Phi_E:=\lambda(E,\cdot)$ the strength
function of $E$. 
We should point out that 
\begin{equation}
\lambda (E,P_\varphi ) 
=\max\left\{\lambda\in{\Bbb R}\colon\ \lambda P_\varphi\le
E\right\}.\label{maxlam}
\end{equation}
That is, $\lambda (E,P_\varphi )P_\varphi\le E$. 
Indeed, from (\ref{lam}) there exists a
sequence $\lambda _i$ such that $\lambda _iP_\varphi\le E$ and $\lambda
_i\to\lambda (E,P_\varphi )$. 
Since $\lambda _i\ab{\elbows{\varphi ,\psi}}^2\le
(E\psi ,\psi)$ for every $\psi$, we have $\lambda
(E,P_\varphi )\ab{\elbows{\varphi ,\psi}}^2  
\le( E\psi ,\psi)$ for
every $\psi$; so (\ref{maxlam}) follows.

\begin{Lem}\label{interpol}
 For any $E\in {\cal E}(\hscript)$ and unit vector $\varphi \in
\hscript$ there exists a unit vector $\psi \in H$ and a $\lambda \in
{\Bbb R}$ such that $\lambda P_\psi \le E$ and $\lambda ( P_\psi
\varphi ,\varphi ) =( E\varphi ,\varphi ) $.
 \end{Lem}

\begin{pf} If $\elbows{E\varphi ,\varphi}=0$, then $\lambda =0$
satisfies the above conditions so suppose\break
$\elbows{E\varphi ,\varphi}\ne 0$. Let $\psi =E\varphi/\|E\varphi\|$ and let
$\lambda =\|E\varphi\|^2/\elbows{E\varphi ,\varphi}$. Then
\begin{equation*}P_\psi\varphi =\elbows{\varphi ,\psi}\psi
  ={{\elbows{E\varphi ,\varphi}}\over{\|E\varphi\|^2}}\,E\varphi
  ={1\over\lambda}E\varphi,
\end{equation*}
so that $\lambda\elbows{P_\psi\varphi ,\varphi}=\elbows{E\varphi ,\varphi}$.
Since $(\psi _1,\varphi _1)\mapsto\elbows{E\psi _1,\varphi _1}$ is a
positive semi-definite sesquilinear form, it follows from Schwarz's
inequality that
\begin{equation*}
|\elbows{E\psi_1 ,\varphi}|^2
  \le\elbows{E\psi _1,\psi _1}\elbows{E\varphi ,\varphi}
\end{equation*}
for any $\psi _1\in H$. Hence, for any $\psi _1\in H$ we have
\begin{eqnarray*}
\lambda\elbows{P_\psi\psi _1,\psi _1}
  &=&\lambda |\elbows{\psi _1,\psi}|^2
  ={\lambda\over{\|E\varphi\|^2}}|\elbows{E\psi _1,\varphi}|^2\\
  &=&{1\over{\elbows{E\varphi ,\varphi}}}\,|
  \elbows{E\psi _1,\varphi}|^2\le\elbows{E\psi _1,\psi _1}.
\end{eqnarray*}
Thus, $\lambda P_\psi\le E$.%$\square$
\end{pf}

\begin{Thm}\label{orderisom}
Let $E,F\in {\cal E}(\hscript)$. Then the following are equivalent:
\newline
$\rm{(i)}$ $\lambda (E,P_{\varphi} ) \le\lambda (F,P_{\varphi} )$ for all
unit vectors $\varphi \in \hscript$;\newline
$\rm{(ii)}$ $E\le F$.
\end{Thm}

\begin{pf} Suppose (i) holds. 
Let $\varphi\in H$ with $\|\varphi\|=1$. Let $\psi$ and
$\lambda$ satisfy the conditions of Lemma \ref{interpol}. Now
$\lambda\le\lambda (E,P_\psi )\le\lambda (F,P_\psi )$. Hence,
$\lambda P_\psi\le\lambda (F,P_\psi )P_\psi\le F$. It follows that
\begin{equation*}
\elbows{E\varphi ,\varphi}=\lambda\elbows{P_\psi\varphi ,\varphi}
  \le\elbows{F\varphi ,\varphi}.
\end{equation*}
Thus, $E\le F$. 
The converse implication is a trivial consequence of the fact that
$E\le F$ entails the implication $\lambda P_\varphi\le E$ $\Rightarrow$
$\lambda P_{\varphi}\le F$. 
%and by symmetry $F\le E$ so that $E=F$.%$\square$
\end{pf}
Thus the map $\Phi:E\mapsto \Phi_E$ is an  order isomorphism from the
set of effects onto the set of strength functions.
This result immediately yields the following.
\begin{Cor}For $E,F\in {\cal E}(\hscript)$, if $\lambda (E,P_{\varphi} )
=\lambda (F,P_{\varphi} )$ for all
unit vectors $\varphi \in \hscript$, then $E= F$.
\end{Cor}
\noindent Hence the map $E\mapsto \Phi_E$ is injective. 
 
\begin{Cor}\label{sup} For any $E\in\escript (H)$ we have
$E=\bigvee\brac{P_\varphi\wedge E\colon\ \|\varphi\|=1}$.
\end{Cor}

\begin{pf}It is easy to show that
$P_\varphi\wedge E=\lambda (E,P_\varphi )P_\varphi$. Now
$P_\varphi\wedge E\le E$ for all $P_\varphi$. Suppose
$F\in\escript (H)$ with $P_\varphi\wedge E\le F$ for all $P_\varphi$.
Then $\lambda (E,P_\varphi )\le\lambda (F,P_\varphi )$ for all unit 
vectors
$\varphi\in H$. Applying Theorem \ref{orderisom}, we have
$E\le F$.
\end{pf}

\section{Properties of the Strength Function}

An element $W\in{\cal E} (\hscript)$ 
is a {\sl weak atom} if for all
$E\in{\cal E} (\hscript)$, $E\le W$ implies 
$E=\lambda W$ for some $\lambda\in [0,1]$.

\begin{Lem}$W$ is a weak atom if and only if
$W=\lambda P_\varphi$ for some $\lambda\in [0,1]$, $\varphi\in \hscript$.
\end{Lem}
\begin{pf}If $E\le\lambda P_\varphi$, then
$\mathrm{ker}(P_\varphi )\subseteq\mathrm{ker}(E)$. Hence,
\begin{equation*}
\mathrm{ran}(E)\subseteq\mathrm{ker}(E)^\perp
  \subseteq\mathrm{ker}(P_\varphi)^\perp
=\mathrm{ran}(P_\varphi)
\end{equation*}
It follows that $E=\lambda _1P_\varphi$ for some
$\lambda _1\le\lambda$ so $\lambda P_\varphi$ is a weak atom.
Conversely, let $W\ne 0$ be a weak atom. Then by 
Corollary \ref{sup}, there exists a $\lambda >0$ and a $P_\varphi$ such
that $\lambda P_\varphi\le W$. Hence,
$\lambda P_\varphi =\lambda _1W$ so
$W=(\lambda /\lambda _1)P_\varphi$.
\end{pf}

Denote the set of weak atoms by $\wscript (\hscript)$. 
The next result follows from
Corollary \ref{sup}. 
\begin{Cor}For any $E\in\escript (\hscript)$ we have
\begin{equation*}
E=\bigvee\brac{W\in\wscript (\hscript)\colon\ W\le E}.
\end{equation*}
\end{Cor}

We have seen that $E\mapsto \Phi_E=\lambda (E,\cdot )$ 
is injective. We now show
that $P_\varphi\mapsto \lambda(\cdot ,P_\varphi )$ is injective. 
\begin{Lem}If
$\lambda (E,P_\varphi )=\lambda (E,P_\psi )$ for every
$E\in\escript (H)$, then $P_\varphi =P_\psi$.
\end{Lem}
\begin{pf}Since $\lambda (P_\psi ,P_\psi )=1$, we have
$\lambda (P_\psi ,P_\varphi )=1$. Hence $P_\varphi\le P_\psi$ and by
symmetry $P_\psi\le P_\varphi$.
\end{pf}

Let $\pscript (\hscript)$ be the set of projections on $\hscript$. 
It is well known that $\pscript (\hscript)$ is the set of extremal 
elements of ${\cal E} (\hscript)$. The next
result shows that $\lambda (\cdot, P_\varphi )$ is a characteristic 
function on $\pscript (\hscript)$.

\begin{Lem}\label{proj} ${\rm (a)}$ $E\in\pscript (\hscript)$ if and only if
$\lambda (E,P_\varphi )\in\brac{0,1}$ for all $P_\varphi$.\newline
${\rm (b)}$ If $P\in\pscript (\hscript)$, 
then $\lambda (P,P_\varphi )=1$
if and only if $\varphi\in\mathrm{ran}(P)$.
\end{Lem}
\begin{pf}(a) Suppose $E\in\pscript (H)$ and
$\lambda (E,P_\varphi )\ne 0$. Then $\lambda P_\varphi\le E$ for some
$\lambda\in (0,1]$. Hence,
$\mathrm{ker}(E)\subseteq\mathrm{ker} (P_\varphi )$ so
$\mathrm{ran}(P_\varphi )\subseteq\mathrm{ran}(E)$. It follows
that $P_\varphi\le E$ so $\lambda (E,P_\varphi )=1$. Conversely,
suppose $\lambda (E,P_\varphi )\in\brac{0,1}$ for all $P_\varphi$.
Then $E\wedge P_\varphi\in\brac{0,P_\varphi}$ for every $P_\varphi$. By
Corollary \ref{sup} we have
\begin{equation*}
E=\bigvee (E\wedge P_\varphi )
  =\bigvee\brac{P_\varphi\colon\ E\wedge P_\varphi =P_\varphi}.
\end{equation*}
But it is well known that the supremum of projections exists and is a
projection. The proof of (b) is obvious.
\end{pf}

The next result shows that $\lambda (\cdot ,P_\varphi )$ is homogeneous,
concave and superadditive.

\begin{Thm}\label{props} ${\rm (a)}$ If $\alpha\in [0,1]$, then
$\lambda (\alpha E,P_\varphi )=\alpha\lambda (E,P_\varphi )$.\newline
${\rm (b)}$ If $\alpha\in [0,1]$, then
\begin{equation*}
\lambda (\alpha E+(1-\alpha )F,P_\varphi )
  \ge\alpha\lambda (E,P_\varphi )+(1-\alpha )\lambda (F,P_\varphi ).
\end{equation*}
${\rm (c)}$ If $E,F\in\escript (\hscript)$ with 
$E+F\in\escript (\hscript)$, then
\begin{equation*}
\lambda (E+F,P_\varphi )
  \ge\lambda (E,P_\varphi )+\lambda (F,P_\varphi ).
\end{equation*}
\end{Thm}
\begin{pf}(a) If $\alpha =0$ the result is clear. If
$\alpha\ne 0$, we have
\begin{equation*}
\begin{split}
\alpha\lambda (E,P_\varphi )&=\sup\brac{\alpha\lambda\in{\real}
  \colon\ \lambda P_\varphi\le E}\\
  &=\sup\brac{\lambda '\in{\real}
  \colon\ \frac{\lambda '}{\alpha}P_\varphi\le E}\\
  &=\sup\brac{\lambda '\in{\real}\colon\ \lambda 'P_\varphi
  \le\alpha E}=\lambda (\alpha E,P_\varphi ).\\
\end{split}
\end{equation*}
(b) Since $\lambda (E,P_\varphi )P_\varphi\le E$ and
$\lambda (F,P_\varphi )P_\varphi\le F$, we have
\begin{equation*}
\sqbrac{\alpha\lambda (E,P_\varphi )
  +(1-\alpha )\lambda (F,P_\varphi )}P_\varphi
  \le\alpha E+(1-\alpha )F
\end{equation*}
and the result follows. The proof of (c) is similar.
\end{pf}

In general, $\lambda (\cdot ,P_\varphi )$ is not additive. For example,
let $P,Q\in\pscript (\hscript)$ with $P\perp Q$ and suppose that
$\varphi\in{\rm ran} (P+Q)$ but $\varphi\notin{\rm ran}(P)$,
$\varphi\notin{\rm ran}(Q)$. Applying Lemma \ref{proj} we have 
 \begin{equation*}
\lambda (P+Q,P_\varphi )=1\ne 0=\lambda (P,P_\varphi ) 
+\lambda (Q,P_\varphi ).
\end{equation*}
We can also use this example to show that 
$\lambda (\cdot ,P_\varphi )$ is not
affine. Applying Theorem \ref{props}(a) 
\begin{equation*}
\lambda\paren{\half P+\half Q,P_\varphi} =\half\lambda (P+Q,P_\varphi )\ne\half%
\lambda (P,P_\varphi ) +\half\lambda (Q,P_\varphi ).
\end{equation*}

It is interesting to study the relation between
the spectrum of $E$, $\sigma (E)$, and the set 
\begin{equation}
\Lambda(E):=
\{\lambda(E,P_\varphi)\colon\varphi\in{\cal{H}},\|\varphi\|=1\}.
\end{equation}
It is easy to see that $\Lambda(E)$ contains the point spectrum 
$\sigma _p(E)$. 
Indeed, if $\lambda\in\sigma _p(E)$ 
then $E\varphi =\lambda\varphi$ for some $\varphi\in \hscript$ 
with $\|\varphi\|=1$. 
Hence, $\lambda =( E\varphi ,\varphi)$. It follows from
the spectral theorem that $\lambda P_\varphi\le E$. 
Since $\lambda^{\prime}P_\varphi\le E$ 
implies that $\lambda ^{\prime}\le( E\varphi,\varphi)=
\lambda$ we have that $\lambda =\lambda (E,P_\varphi )$. We thus
obtain the following. 
\begin{Lem}\label{eigv1}$\rm{(a)}$ $\lambda(E,P_\varphi)\le
(\varphi,E\varphi)$.\newline
$\rm{(b)}$ If $E\varphi =\lambda\varphi$, $\|\varphi\|=1$, then
$\lambda =\lambda (E,P_\varphi )=\elbows{E\varphi ,\varphi}$.
\end{Lem}

More information on the connection between $\sigma(E)$ and $\Lambda(E)$ 
can be obtained once the explicit form of the strength function is
determined. This will be established in
the next section.

\section{Explicit Form of the Strength Function}

Let $E\in\escript(\hscript)$ 
and $X\mapsto P^E(X)$ its spectral measure. For $\varepsilon \in (0,1)$, we
let $P_\varepsilon :=P^E\left( \left[ \varepsilon ,1\right] \right) $ 
and $E_\varepsilon :=EP_\varepsilon $. 
By $E^{-1/2}$ we denote the inverse of the injective map 
$E^{1/2}|_{{\cal H}_{0}}$, 
where ${\cal H}_{0}$ is the closure of the range of $E^{1/2}$.

\begin{Thm}\label{root}
Let $E$ be an effect, $P_\varphi $ a one-dimensional projection. Then 
\begin{equation}
\exists \lambda >0:\lambda P_\varphi \le E\quad \Leftrightarrow \quad
\varphi \in {\rm ran}\left( E^{1/2}\right) .  
\label{lowbd}
\end{equation}
\end{Thm}

\begin{pf} The statement has appeared rather implictly, and 
was proved in \cite{Bus84} using techniques applicable in 
separable Hilbert spaces. Here we present a more transparent 
proof for general Hilbert spaces.
It is easy to see that $\varphi \in \mathrm{ran}(E^{1/2})$ is sufficient for 
$\lambda P_\varphi \le E$ to hold with
some positive $\lambda $. In fact let $\xi $ be the unique element in 
${\mathcal{H}}_{0}$ such that 
$\varphi =E^{1/2}\xi $. Let $\lambda =\left\| \xi \right\| ^{-2}=\left\|
E^{-1/2}\varphi \right\| ^{-2}$. Then for any $\psi \in \mathcal{H}$, we have
\begin{eqnarray*}
\left( \psi ,\lambda P_\varphi \psi \right) &=&
\lambda \left( E^{1/2}\psi ,\xi
\right) \left( \xi ,E^{1/2}\psi \right)\\ 
&=& \lambda \left\| \xi \right\|
^2\left( E^{1/2}\psi ,P_\xi E^{1/2}\psi \right) \le \lambda \left\| \xi
\right\| ^2\left( \psi ,E\psi \right) .
\end{eqnarray*}
With the above choice of $\lambda $, this yields $\lambda P_\varphi
\le E$. 

The proof of the converse implication is somewhat more involved. Still it is
not hard to show that $\lambda P_\varphi \le E$ with $\lambda >0$
necessitates $\varphi \in {\mathcal{H}}_{0}$. Suppose $\varphi \notin 
{\mathcal{H}}_{0}$. 
There exist a unique decomposition $\varphi =\varphi _0+\varphi _1$,
with $\varphi _0\in {\mathcal{H}}_{0}$, $\varphi _1\in \ker \left(
E^{1/2}\right) ={\mathcal{H}}_{0}^{\perp }$, and $\varphi _1\neq 0$. 
Now suppose 
$\lambda P_\varphi \le E$. For $\psi =\varphi _1/||\varphi _1||$, this
implies $\lambda |(\psi ,\varphi )|^2=\lambda ||\varphi _1||^2\le (\psi
,E\psi )=0$, and so $\lambda =0$. 

This observation makes it possible to assume in the remaining part of the
proof that all vectors involved are elements of ${\mathcal{H}}_{0}$. 
We assume
again that there exists a $\lambda >0$ such that 
$\lambda P_\varphi \le E$. 
We have to show that this implies $\varphi \in \mathrm{ran}(E^{1/2})$. 
The assumption means the following:
\begin{equation}
\exists \lambda >0\quad \forall \psi \in 
{\mathcal{H}}_0\quad 0\le 
\| E^{1/2}\psi \| ^2-\lambda \left\| P_\varphi \psi
\right\| ^2.  \label{lb1}
\end{equation}
Note that for $\psi $ collinear with $\varphi $ the inequality only yields 
\begin{equation}
\lambda \le (\varphi ,E\varphi ),  \label{lb2}
\end{equation}
which confirms the statement of Lemma \ref{eigv1}(a).
So we can assume $\psi\not\in [\varphi]$, and condition 
(\ref{lb1}) is equivalent to:
\begin{equation}
\exists \lambda >0\quad \forall \psi 
\in {\mathcal{H}}_0\backslash [\varphi
]\quad 0\le \left| \left| E^{1/2}\psi \right| \right| ^2-\lambda \left|
\left| P_\varphi \psi \right| \right| ^2.  \label{lb3}
\end{equation}
Any such $\psi $ can be taken to be of the form 
$\psi =\xi +xe^{it}\varphi ,$
where $\xi $ is a unit vector orthogonal to $\varphi $, 
$x>0$, $0\le t$. Substituting $\psi $ with this expression in  
(\ref{lb3}) gives the equivalent statement
\begin{equation}
\begin{split}
\exists \lambda >0\quad \forall \xi \in 
{\mathcal{H}}_0\cap [\varphi ]^{\perp
}\quad \forall x,t\quad 0 &\le \left\| E^{1/2}\xi \right\|^2
+x^2\left| \left| E^{1/2}\varphi \right| \right| ^2\\ 
&+ 2x{\rm{Re}}\left\{
e^{it}\left( E^{1/2}\varphi ,E^{1/2}\xi \right) \right\} -\lambda x^2,
\end{split}
\end{equation}
or equivalently, 
\begin{equation}\begin{split}
\exists\lambda>0\ 
\forall \xi \in {\mathcal{H}}_0\cap [\varphi ]^{\perp }\quad 
\forall x,t\quad
0< \lambda &\le\left\| E^{1/2}\varphi \right\|  ^2+x^{-2}\left\| 
E^{1/2}\xi \right\|  ^2\\
&+2x^{-1}{\rm{Re}}\left\{ e^{it}\left(
E^{1/2}\varphi ,E^{1/2}\xi \right) \right\} .
\end{split}
\end{equation}
Minimising the right hand side of  with respect to $t$ gives the equivalent
condition 
\begin{equation*}\begin{split}
\exists\lambda>0\ \forall \xi \in {\mathcal{H}}_0\cap [\varphi ]^{\perp }\quad 
\forall x>0\quad
0<\lambda &\le\left| \left| E^{1/2}\varphi \right| \right| ^2
+x^{-2}\left| \left|
E^{1/2}\xi \right| \right| ^2\\
&-2x^{-1}\left| \left( E^{1/2}\varphi
,E^{1/2}\xi \right) \right| .
\end{split}
\end{equation*}
Finally, minimising this expression with respect to $y=x^{-1}$ yields:
\begin{equation}
\exists\lambda>0\ \forall \xi \in [\varphi ]^{\perp }\quad 
0<\lambda\le\left| \left| E^{1/2}\varphi
\right| \right| ^2-
\left| \left(E^{1/2}\varphi ,E^{1/2}\xi \right) \right| ^2\,
\left\| E^{1/2}\xi \right\|^{-2}.  \label{lb4}
\end{equation}
Thus we have established the equivalence of condition 
(\ref{lb1}) with  (\ref{lb4}). 
But this last inequality states that $E^{1/2}\varphi $ must have a
finite positive distance from the closure of the subspace 
$E^{1/2}([\varphi]^{\perp })$. 
Let $Q$ denote the associated projector. Then (\ref{lb1}) is
seen to be equivalent to the following:
\begin{equation}
0<\left\| (\mathrm{I}-Q)E^{1/2}\varphi \right\| ^2,  \label{lowbd2}
\end{equation}
this number being an upper bound for $\lambda $. 
We just noted that the condition (\ref{lowbd2}) is equivalent to the 
following:
\begin{equation}
E^{1/2}\varphi \notin \overline{E^{1/2}([\varphi ]^{\perp })}.  
\label{notin}
\end{equation}
Finally we show the equivalence of (\ref{notin}) with
\[
\varphi \in \mathrm{ran}(E^{1/2}).
\]

Let $\varphi =E^{1/2}\xi $. Then for all $\eta \in {\mathcal{H}}_{0},$ $\eta
\perp \varphi $, we have $0=(\varphi ,\eta )=(\xi ,E^{1/2}\eta )$, and so 
$(I-Q)\xi =\xi $. It follows that 
\[
1=(\varphi ,\varphi )=(\xi ,E^{1/2}\varphi )=((\mathrm{I}-Q)\xi
,E^{1/2}\varphi )=(\xi ,(\mathrm{I}-Q)E^{1/2}\varphi ),
\]
thus $(\mathrm{I}-Q)E^{1/2}\varphi \neq 0,$ that is (\ref{notin}) follows.

Conversely, assume $\varphi \notin \mathrm{ran}(E^{1/2})$. Let $\eta \in 
{\mathcal{H}}_0$, $\eta \perp \overline{E^{1/2}([\varphi ]^{\perp })}$. Thus,
for all $\psi \in {\mathcal{H}}_{0},$ $\psi \perp \varphi ,$ we have $0=(\eta
,E^{1/2}\psi )=(E^{1/2}\eta ,\psi ),$ and this implies $E^{1/2}\eta =\alpha
\varphi .$ Now $\varphi \notin \mathrm{ran}(E^{1/2})$ implies $\alpha =0,$
therefore  $E^{1/2}\eta =0$ and finally $\eta =0.$ Thus, $E^{1/2}([\varphi
]^{\perp })$ is dense in ${\mathcal{H}}_{0},$ and so $E^{1/2}\varphi \in 
\overline{E^{1/2}([\varphi ]^{\perp })}$. 
\end{pf}

Theorem \ref{root} allows us to determine the explicit form of
$\lambda(E,P_\varphi)$ and along with this an alternative proof of the
injectivity of $E\mapsto \Phi_E$. 

\begin{Thm}\label{form}
The numbers $\lambda(E,P_\varphi)$ are of the following form: 
\begin{equation}
 \lambda(E,P_\varphi)=\begin{cases}
\left\| E^{-1/2}\varphi \right\| ^{-2},
&\text{if $\varphi \in{\rm{ran}}\left( E^{1/2}\right)$;}  \\ 
\qquad 0, &\text{else.}
\end{cases}
\label{form1}
\end{equation}
The map $E\mapsto \Phi_E=\lambda(E,\cdot)$ is injective.
\end{Thm}

\begin{pf} Let $\varphi \in \mathcal{H},$ $\left\|
\varphi \right\| =1$. 
If $\varphi \notin \mathrm{ran}\left( E^{1/2}\right)$,
Theorem \ref{root} 
implies that %$\varphi _E\left( P_\varphi \right) 
$\lambda(E,P_\varphi)=0$. Next assume 
$\varphi \in \mathrm{ran}\left( E^{1/2}\right) $. Due to Theorem
\ref{root} we have
the following: 
\begin{equation}
\lambda P_\varphi \le E,  \label{ineq1}
\end{equation}
that is, 
\begin{equation}
\lambda \left| \left( \psi ,\varphi \right) \right| ^2\le
\| E^{1/2}\psi \| ^2,\quad \forall \psi \in \mathcal{H}.
\label{ineq2}
\end{equation}
This in turn implies: 
\begin{equation}
\lambda \big| ( E^{1/2}\psi ,\xi) \big| ^2\le
\| E^{1/2}\psi \| ^2,\quad \xi =E^{-1/2}\varphi ,\quad \forall
\psi \in {\mathcal{H}}.  \label{ineq3}
\end{equation}
The Cauchy-Schwarz inequality gives an upper bound for the left hand side: 
\begin{equation*}
\big| ( E^{1/2}\psi ,\xi ) \big| ^2\le \|
E^{1/2}\psi \| ^2\| \xi \| ^2,
\end{equation*}
and therefore Eq. (\ref{ineq3}) is implied by 
\begin{equation}
\lambda \le \left\| \xi \right\| ^{-2}=\left\| E^{-1/2}\varphi \right\|
^{-2}.  \label{supm}
\end{equation}
We show that Eq.~ (\ref{supm}) is also necessary for 
Eq.~ (\ref{ineq3}). For 
$\varepsilon \in (0,1),$ let $\psi _\varepsilon =E_\varepsilon ^{-1/2}\xi $.
This is well defined since $E_\varepsilon ^{1/2}$ is invertible on 
${\mathcal{H}}_{0}$. 
We have $E^{1/2}\psi _\varepsilon =E^{1/2}E_\varepsilon ^{-1/2}\xi
=P_\varepsilon \xi .$ Thus, putting $\psi =\psi _\varepsilon $, Eq. (\ref
{ineq3}) implies 
\begin{equation*}
\lambda \left| \left( P_\varepsilon \xi ,\xi \right) \right|
^2=\lambda \left\| P_\varepsilon \xi \right\| ^4\le \left\| P_\varepsilon
\xi \right\| ^2.
\end{equation*}
For sufficiently small $\varepsilon $ we have $P_\varepsilon \xi \ne 0$, and
so for $\varepsilon \rightarrow 0$, we conclude that Eq.~ (\ref{supm}) must
hold. Therefore the number 
$\| E^{-1/2}\varphi \| ^{-2}=\lambda(E,P_\varphi)$.

Next we show the injectivity of the map $E\mapsto \Phi_E=\lambda(E,\cdot)$. 
For two effects $E,F$, assume that $\Phi _E=\Phi _F$. 
Then we have: $\varphi \notin \mathrm{%
ran}\left( E^{1/2}\right) $ iff $\Phi _E\left( P_\varphi \right) =0$, iff $%
\Phi _F\left( P_\varphi \right) =0$, iff $\varphi \notin \mathrm{ran}\left(
F^{1/2}\right) $. It follows that $\mathrm{ran}\left( E^{1/2}\right) =%
\mathrm{ran}\left( F^{1/2}\right) =:\mathcal{R}$. Now, for $\varphi \in 
\mathcal{R}$, there are unique elements $\xi ,\eta \in {\mathcal{H}}_{0}$ such
that  $\varphi =E^{1/2}\xi =F^{1/2}\eta $. Then we have: 
\begin{equation*}
\| E^{-1/2}\varphi \| ^{-2}=\| F^{-1/2}\varphi \|
^{-2}\quad \forall \varphi \in \mathcal{R}
\end{equation*}
if and only if
\begin{equation*}
\| E^{-1/2}F^{1/2}\eta \| =\| \eta \| \quad \forall
\eta \in {\mathcal{H}}_{0}
\end{equation*}
if and only if
\begin{equation*}
\| F^{-1/2}E^{1/2}\xi \| =\| \xi \| \quad \forall \xi
\in {\mathcal{H}}_{0}.
\end{equation*}
This shows that the operators $A:=E^{-1/2}F^{1/2}$ and $B:=F^{-1/2}E^{1/2}$
are bounded and indeed unitary on ${\mathcal{H}}_0$. 
Moreover, 
$AB=BA=\mathrm{I}_{{\mathcal{H}}_{0}}$. 
So $B=A^{*}$, and this is the extension of the densely
defined and bounded operator $F^{1/2}E^{-1/2}:{\mathcal{R}}\rightarrow 
{\mathcal{H}}_{0}$. 
Thus, $F^{-1/2}E^{1/2}=F^{1/2}E^{-1/2}$. Multiplying this
from the left with $F^{1/2}$ and from the right with $E^{1/2}$ finally gives 
$E=F$. 
\end{pf}

We are now in a position to prove the converse to Lemma \ref{eigv1}(b).

\begin{Prop}\label{eigv2} 
\[\lambda (E,P_\varphi )= (E\varphi,\varphi)\quad\Leftrightarrow\quad
E\varphi =\lambda(E,P_\varphi)\varphi.
\] 
\end{Prop}

\begin{pf}
In view of
Lemma \ref{eigv1}(b), it remains to show that the equation
$\lambda(E,P_\varphi)=(\varphi,E\varphi)$ implies
$E\varphi=(\varphi,E\varphi)\varphi$. 

In case $\varphi\not\in{\mathrm{ran}}(E^{1/2})$, we have
$0=\lambda(E,P_\varphi)$, so that the equation implies $E\varphi=0$.

Now let $\varphi\in {\mathrm{ran}}(E^{1/2})$. Theorem \ref{form} gives
for $\xi=E^{-1/2}\varphi$: 
$1=\|\xi\|^2\,\|E\xi\|^2$, and combining this with the Cauchy-Schwarz
inequality, we obtain
\[
1=\|\xi\|\,\|E\xi\|\ge (\xi,E\xi)=\|E^{1/2}\xi\|^2=\|\varphi\|^2=1.
\]
But equality holds only if $E\xi=\alpha\xi$, therefore
$E\varphi=\alpha\varphi$.
\end{pf}

Finally we establish the precise relation between $\sigma(E)$ and
$\Lambda(E)$. 

\begin{Thm}
Let $[a,b]$ be the convex hull of $\sigma (E),$  so that $a=\min (\sigma
(E))=1-||{\mathrm{I}}-E||,$ $b=\max (\sigma (E))$. Then:%\newline
\begin{itemize}
\item[$\rm{(a)}$] If $a>0,$ then $a\in \Lambda (E)$ 
$\Leftrightarrow a\in \sigma _p(E).$ 
%\newline
\item[$\rm{(b)}$] 
$b\in \Lambda (E)\Leftrightarrow b\in \sigma _p(E)$.%\newline
\item[$\rm{(c)}$] If $a>0,$ then $(a,b)\subseteq 
\Lambda (E)=\{(\varphi ,E\varphi
)\colon\varphi \in {\mathcal{H}},||\varphi ||=1\}\subseteq [a,b]$.%\newline
\item[$\rm{(d)}$] If $a=0$ is an isolated eigenvalue of $E$, 
then $\sigma (E)=\{0\}\cup
[a_0,b]$,\newline 
and $\{0\}\cup (a_0,b)\subseteq \Lambda (E)\subseteq \{0\}\cup
[a_0,b],$ where $a_0=\min (\sigma (E)\backslash \{0\})$.%\newline
\item[$\rm{(e)}$] If $a=0$ is an accumulation point in $\sigma (E)$, then 
$[0,b)\subseteq \Lambda (E)\subseteq [0,b]$.
\end{itemize}
\end{Thm}

\begin{pf}
(a) If $a$ is an eigenvalue of $E$, there exists a unit vector 
$\varphi $ such that $E\varphi =a\varphi $. 
By Lemma \ref{eigv1}, $a=\lambda(E,P_\varphi )\in \Lambda (E)$. 
Conversely, let $a=\lambda (E,P_\varphi)=
||E^{-1/2}\varphi ||^{-2}$ for some unit vector $\varphi 
\in {\mathcal{H}}=\mathrm{ran}(E^{1/2})$. 
By virtue of the spectral theorem, 
$a=||E^{-1/2}||^{-2}$. Therefore, taking into account the fact  
that $||E^{-1/2}||^2\mathrm{I}-E^{-1}\ge 0$, we have 
\[
0=||E^{-1/2}||^2-||E^{-1/2}\varphi ||^2=\left\|\left 
(||E^{-1/2}||^2\mathrm{I}%
-E^{-1}\right)^{1/2}\varphi \right\| ^2,
\]
which implies $E^{-1}\varphi =a^{-1}\varphi$, and so 
$E\varphi =a\varphi $.

The proof of (b)  follows by a similar method.

(c) Note that $\mathrm{ran}(E^{1/2})={\mathcal{H}}$. 
Hence
$\varphi $, $\lambda (E,P_\varphi )=
||E^{-1/2}\varphi ||^{-2}\allowbreak\in [a,b]$ for all $\varphi$ since 
$\sigma (E^{-1/2})\subseteq [b^{-1/2},a^{-1/2}]$. Therefore, $\Lambda
(E)\subseteq [a,b]$. Next let $\alpha ,\beta \in [0,1]$ 
such that $a<\alpha<\beta <b$. 
Let $\varphi _\alpha ,$ $\varphi _\beta $ be unit vectors such that 
$P^E([a,\alpha]) \varphi _\alpha =\varphi _\alpha ,$ 
$P^E([\beta,b]) \varphi _\beta
=\varphi _\beta .$ Then it is easy to see that $\lambda (E,P_{\varphi
_\alpha })\in [a,\alpha ],$ 
$\lambda (E,P_{\varphi _\beta })\in [\beta ,b].$
Define the unit vector $\varphi =\sqrt{w}\varphi _\alpha +
\sqrt{1-w}\varphi_\beta ,$ where $w\in [0,1].$ We compute
\begin{eqnarray*}
\lambda (E,P_\varphi )^{-1} &=&||E^{-1/2}\varphi ||^2=w||E^{-1/2}\varphi
_\alpha ||^2+(1-w)||E^{-1/2}\varphi _\beta ||^2 \\
&=&w\lambda (E,P_{\varphi _\alpha })+(1-w)\lambda (E,P_{\varphi _\beta }).
\end{eqnarray*}
Varying $w$ between 0 and 1, we see that $\lambda (E,P_\varphi )$ assumes
all values between $\lambda (E,P_{\varphi _\alpha })$ and $\lambda
(E,P_{\varphi _\beta }),$ and so $(\alpha ,\beta )\subseteq \Lambda (E)$.
Since $\alpha ,\beta \in (a,b)$ are arbitrary, it follows that 
$(a,b)\subseteq \Lambda (E).$ 
The equation $\Lambda (E)=\{(\varphi ,E\varphi)\colon\varphi 
\in {\mathcal{H}},||\varphi ||=1\}$ follows easily from the spectral
theorem.

(d) Let $a=0$ be an isolated eigenvalue of $E$. Then $\mathrm{ran}%
(E^{1/2})=\mathcal{H}_0$ is a closed subspace of $\mathcal{H},$ and for all
unit vectors $\varphi \in {\mathcal{H}}_0,$ $\lambda (E,P_\varphi )=\lambda
(E|_{{\mathcal{H}}_{0}},P_\varphi )=||E^{-1/2}\varphi ||^{-2}.$ So 
$(a_0,b)\subseteq \Lambda (E|_{{\mathcal{H}}_{0}})\subset \Lambda (E)$. 
For all
unit vectors $\varphi $ not in ${\mathcal{H}}_0,$ 
$\lambda (E,P_\varphi )=0\in
\Lambda (E).$ It is easy to see that $a_0$ is a lower bound of $\Lambda
(E)\backslash \{0\}$ and $b$ is an upper bound. This proves (d).

(e) Let $a=0\in \sigma _c(E),$ the continuous spectrum of $E$. There
exists a decreasing sequence $\alpha _n\in \sigma (E)\cap (0,b]$ with limit
0. For $E_n=EP^E([\alpha _n,b])$ and any unit vector $\varphi \in
P^E([\alpha _n,b]),$ one finds $\lambda (E,P_\varphi )=\lambda
(E_n,P_\varphi ).$ Therefore, $(\alpha _n,b)\subseteq \Lambda (E_n)\subseteq
\Lambda (E)$ for all $n\in {\Bbb N},$ and so $(0,b)\subseteq \Lambda (E)\subseteq
[0,b].$
\end{pf}


\begin{thebibliography}{9}
\bibitem{Lud} G.~ Ludwig, {\em Foundations of Quantum Mechanics I},
          Springer Verlag, Berlin, 1983.
\bibitem{Bus84} P.~ Busch, J.~ Math.~ Phys.~ 25 (1984) 
1794--1797. 
\end{thebibliography}
\end{document}